\documentclass[useAMS,usenatbib]{mn2e}
\usepackage{graphics,graphicx}
\usepackage{aas_macros}
\usepackage{amssymb}
\usepackage{latexsym}
\usepackage{color}
\usepackage{epstopdf}
\usepackage{marvosym}

\begin{document}

\title[Collisional excitation of HC$_3$N by H$_2$]{Collisional
  excitation of HC$_3$N by para- and ortho-H$_2$} \author[Alexandre
  Faure, Fran\c{c}ois Lique and Laurent Wiesenfeld]{Alexandre
  Faure$^{1, 2}$\thanks{E-mail:
    alexandre.faure@univ-grenoble-alpes.fr}, Fran\c{c}ois Lique$^3$\thanks{E-mail: francois.lique@univ-lehavre.fr} and Laurent
  Wiesenfeld$^{1, 2}$\thanks{E-mail:
    laurent.wiesenfeld@univ-grenoble-alpes.fr} \\ $^1$
  Univ. Grenoble Alpes, IPAG, F-38000 Grenoble, France\\ $^2$ CNRS, IPAG, F-38000 Grenoble, France \\ $^3$ LOMC-UMR 6294, CNRS-Universit\'e du Havre, 25 rue
  Philippe Lebon, BP 1123 -- 76063 Le Havre Cedex, France}

\date{Accepted 2016 May 11. Received 2016 May 10; in original form 2016 May 04}

\pagerange{\pageref{firstpage}--\pageref{lastpage}} \pubyear{2016}

\maketitle

\label{firstpage}

\begin{abstract}

New calculations for rotational excitation of cyanoacetylene by
collisions with hydrogen molecules are performed to include the lowest
38 rotational levels of HC$_3$N and kinetic temperatures to
300~K. Calculations are based on the interaction potential of Wernli
et al. {\it A\&A}, 464, 1147 (2007) whose accuracy is checked against
spectroscopic measurements of the HC$_3$N--H$_2$ complex. The quantum
coupled-channel approach is employed and complemented by
quasi-classical trajectory calculations. Rate coefficients for
ortho-H$_2$ are provided for the first time. Hyperfine resolved rate
coefficients are also deduced. Collisional propensity rules are
discussed and comparisons between quantum and classical rate
coefficients are presented. This collisional data should prove useful
in interpreting HC$_3$N observations in the cold and warm ISM, as well
as in protoplanetary disks.

\end{abstract}


\section{Introduction}

Cyanopolyyne molecules, with general formula HC$_{2n+1}$N ($n\geq 1$),
have been detected in a great variery of astronomical environments,
from the solar system to extragalactic sources. The simplest one,
HC$_3$N (cyanoacetylene) is the most abundant of the family. It has
been first detected towards the giant galactic molecular cloud Sgr~B2
by \citet{turner71} and in comet Hale-Bopp by
\citet{bockelee00}. Because of its low rotational constant and large
dipole moment, HC$_3$N is considered as a very good thermometer and
barometer in the ISM. It has been detected in the ground level and in
excited vibrational levels, thanks to the presence of low-lying
bending modes (below 1000~cm$^{-1}$). Cyanoacetylene is also an
abondant nitrogen bearing species and it has been proposed as a
precursor of prebiotic molecules such as cytosine, one of the four
nitrogen bases found in DNA and RNA \citep[e.g][]{orgel02}.

In most astrophysical regions where HC$_3$N is observed, the low
frequency of collisions (due to low density) cannot maintain local
thermodynamical equilibrium (LTE). This is best exemplified by the
natural occurrence of the maser phenomenon (microwave amplification by
stimulated emission of radiation) in the $1\rightarrow 0$ transition
at 9.1~GHz, as first observed towards Sgr~B2 by \citet{morris76}. The
interpretation of HC$_3$N emission spectra, in terms of density,
temperature and column density, thus requires to solve the equations
of statistical equilibrium, which in turn necessitates a good
knowledge of the collisional rate coefficients. In the cold and dense
ISM, the most abundant colliders are hydrogen molecules and helium
atoms. In warmer regions, such as photodissociation regions (PDRs) or
comets, electron collisions may also play an important role, due to
the large HC$_3$N dipole (3.7~D) \citep[see e.g.][]{gratier13}, as
well as hydrogen atoms.
 
The collisional excitation of HC$_3$N by He and H$_2$ was first
studied by \citet{morris76} using the ``hard-ellipsoid'' classical
approximation. This pioneering work was soon followed by the first
determination of the HC$_3$N-He potential energy surface (PES), using
an approximate ``electron gas'' model, combined with Monte-Carlo
quasi-classical trajectory (QCT) calculations
\citep{green78}. Classical mechanics was employed instead of quantum
mechanics due to the difficulty of expanding the PES in terms of
Legendre polynomials. Indeed, for such extremely anisotropic
interactions (HC$_3$N is $\sim$ nine bohr long), the Legendre
polynomial expansion is very slowly convergent and subject to severe
numerical problems, as first discussed by \cite{chapman77}. These
problems were circumvented by a novel approach proposed by
\cite{wernli07} who derived the first PES for HC$_3$N--H$_2$
(hereafter denoted as W07). These authors were thus able to perform
both classical and quantum calculations for HC$_3$N colliding with
para-H$_2$($j_2=0$, where $j_2$ is the H$_2$ angular momentum). They
provided rate coefficients for the lowest 51 rotational levels of
HC$_3$N ($j_1=0-50$, where $j_1$ is the HC$_3$N angular momentum) and
kinetic temperatures in the range 10-100~K. Quantum results were
however restricted to the lowest 16 levels, due to the high
computational cost of these calculations. In addition, numerical
errors in the implementation of the PES routine were found to
introduce small inaccuracies ($\lesssim$20\%) in the quantum
calculations, as reported by \citet{wernli07b}.

In the present work, the W07 PES is first checked against recent
high-resolution spectroscopic measurements of the HC$_3$N--H$_2$
complex. It is then employed to derive new rate coefficients for
HC$_3$N colliding with para-H$_2$($j_2=0$) and, for the first time,
with ortho-H$_2$($j_2=1$). The quantum coupled-channel approach is
employed for the lowest 31 rotational levels of HC$_3$N ($j_1=0-30$)
while the QCT approach is employed for higher levels, up to $j_1=37$
which is the highest level below the first excited bending mode
$\nu_7$ at 222~cm$^{-1}$. Hyperfine selective collisional cross
sections are also deduced using the almost exact (but computationally
demanding) recoupling method and the infinite-order-sudden (IOS)
approximation. In Section~2, the spectra of the HC$_3$N--H$_2$ complex
is computed and it is compared to experimental data. In Section~3,
scattering calculations are described and the procedures used to
derive cross sections and rate coefficients are briefly
introduced. Results are presented in Section~4. We discuss in
particular the propensity rules and a comparison between quantum and
classical rate coefficients is made. Conclusions are summarized in
Section~5.

\section{Interaction potential and bound-states}

All bound-state and scattering calculations presented in the following
sections were performed with the W07 PES. This PES is briefly
described below. The lowest bound-state rovibrational energy levels
supported by the PES are computed and the corresponding HC$_3$N--H$_2$
transitions frequencies are compared to the recent experimental
results of \cite{michaud11}.

\subsection{Potential energy surface}

The W07 PES was computed at the coupled-cluster with single, double,
and perturbative triple excitations [CCSD(T)] level of theory with an
augmented correlation-consistent triple zeta [aug-cc-pVTZ] basis set
for HC$_3$N and quadruple zeta [aug-cc-pVQZ] for H$_2$, as described
in \citet{wernli07}. The molecules were both treated as rigid
rotors. Owing to the computational cost of the calculations (HC$_3$N
is a 26 electron system), the angular sampling of the PES was
performed by using only five independent H$_2$ orientations. The
resulting five PES were interpolated using a bicubic spline function
and the full HC$_3$N--H$_2$ interaction potential was reconstructed
analytically \citep{wernli06}. Due to the steric hindrance caused by
the HC$_3$N rod, a standard spherical harmonics expansion of the full
PES, as required by quantum calculations, was not possible (see
\cite{chapman77}). This problem was solved by \citet{wernli07} using a
novel approach called ``regularization''. This technique consists in
scaling the PES when it is higher than a prescribed threshold so that
the potential smoothyl saturates to a limiting value in the highly
repulsive regions of the PES. The ``regularized'' PES thus retain only
the low energy content of the original PES but it is accurately
adapted to low collisional energies, in practice lower than $\sim
1000$~cm$^{-1}$ here. The final spherical harmonics fit includes terms
up to $l_1=24$ and $l_2=2$, resulting in 97 angular functions. Full
details can be found in \cite{wernli06} and \cite{wernli07}.

The equilibrium structure of the HC$_3$N--H$_2$ complex is linear with
the nitrogen of HC$_3$N pointing towards H$_2$ at an intermolecular
separation of 9.58~bohr. The corresponding well depth is
-193.7~cm$^{-1}$, which compares well with the PES of Michaud et
al. (2011) (hereafter denoted as M11), also computed at the CCSD(T)
level, whose grid minimum is at -188.8~cm$^{-1}$. When averaged over
H$_2$ orientation (corresponding to para-H$_2$ in its ground state
$j_2=0$), the W07 PES has a shallower well of -111.2~cm$^{-1}$, also
in good agreement with the M11 averaged PES minimum of
-112.6~cm$^{-1}$. It should be noted, however, that the calculations
of Michaud et al. (2011) were performed for only three orientations of
the hydrogen molecule within the complex and no global fit of the PES
was provided by the authors.

\subsection{Bound-states}

A spectroscopic and theoretical study of the HC$_3$N--H$_2$ dimer was
reported by Michaud et al. (2011). The bound-state energies of the
complex were computed by these authors using scaled and unscaled
versions of their PES. These calculations were performed in
two-dimensions (2D) using the Lanczos iterative procedure applied to
the averaged M11 PES for para-H$_2$ and to the three PES
(corresponding to the three hydrogen orientation) separately for
ortho-H$_2$. The results were compared with high-resolution microwave
spectroscopy measurements. Theoretical and experimental transition
frequencies were shown to be in good agreement, within a few percent
or better (see below).

We have also computed the bound-state energies supported by the W07
PES using the coupled-channel approach, as implemented in the
\texttt{BOUND} program \citep{hutson94}. The full spherical harmonics
expansion of the W07 PES was employed with 97 functions. The coupled
equations were solved using the diabatic modified log-derivative
method. These calculations were performed in four-dimensions (4D) for
both para- and ortho-H$_2$. Both molecules were taken as rigid rotors
with the rotational constants $B_0$=59.322~cm$^{-1}$ for H$_2$ and
$B_0$=0.151739~cm$^{-1}$ for HC$_3$N. A total of 46 rotational states
(i.e. up to $j_1=45$) were included in the HC$_3$N basis set while the
two lowest rotational states of para-H$_2$ ($j_2=0, 2$) and
ortho-H$_2$ ($j_2=1, 3$) were considered. The calculations were
performed with a propagator step size of 0.01~bohr and the other
propogation parameters were taken as the default \texttt{BOUND}
values.

The transition frequencies in the HC$_3$N--H$_2$ complex were deduced
from the computed bound-state energies. They are compared to the
spectroscopic and theoretical data of Michaud et al. (2011) in
Tables~1 and 2 below. The assignment of the (approximate) quantum
numbers $J_{K_aK_c}$ was performed by Michaud et al. (2011). These
authors reported a set of experimental line frequencies including the
splitting into several hyperfine components due to the nuclear spin of
the nitrogen atom ($I=1$). In the Tables below, the theoretical line
frequencies are compared to the unsplit experimental frequencies of
Michaud et al. (2011). The agreement between the present values and
the experimental data is within 0.5\% (i.e. better than
0.01~cm$^{-1}$) for HC$_3$N--para-H$_2$ and within a few percent
(better than 0.1~cm$^{-1}$) for HC$_3$N--ortho-H$_2$. The
corresponding average differences are 0.32\% and 4.2\%,
respectively. This very good agreement is similar, although slightly
better, than the (nonscaled) theoretical results of Michaud et
al. (2011). This was expected since the level of {\it ab initio}
theory is similar and the improvement is likely due to the 4D
approach. We note that Michaud et al. (2011) were able to obtain an
even better agreement by applying simple scaling techniques (radial
shifts) to the M11 PES.

\begin{table*}
\caption{Transition frequencies (in MHz) in the HC$_3$N--para-H$_2$
  complex from experiment, from the calculations of Michaud et
  al. (2011) and from the present calculations. Deviations from the
  experiment are also given in percent.}
\begin{center}
\begin{tabular}{cccccc}
\hline $J_{K_aK_c}'-J_{K_aK_c}''$ & experiment & Michaud et al. &
error (\%) & present & error (\%) \\
$1_{01}-0_{00}$ & 8295.5435  &  8240.0   & -0.7   & 8268.8  & -0.3  \\
$2_{12}-1_{11}$ & 15601.6795 &  15497.7  & -0.7   & 15529.2 & -0.5  \\
$1_{10}-1_{01}$ &            &  16248.4  &        & 16139.1 &       \\
$2_{02}-1_{01}$ & 16549.1211 &  16436.5  & -0.7   & 16514.4 & -0.2  \\
$2_{11}-1_{10}$ & 17521.8267 &  17418.2  & -0.6   & 17510.4 & -0.07 \\
$3_{13}-2_{12}$ & 23378.90   &  23195.3  & -0.8   & 23283.7 & -0.4  \\
$1_{11}-0_{00}$ &            &  23518.9  &        & 23429.5 &       \\
$3_{03}-2_{02}$ & 24718.7675 &  24545.3  & -0.7   & 24656.7 & -0.3  \\
$3_{12}-2_{11}$ &            &  26100.6  &        & 26242.8 &              \\ \hline
\end{tabular}
\end{center}
\label{table:1}
\end{table*}

\begin{table*}
\caption{Transition frequencies (in MHz) in the HC$_3$N--ortho-H$_2$
  complex from experiment, from the calculations of Michaud et
  al. (2011) and from the present calculations. Deviations from the
  experiment are also given in percent.}
\begin{center}
\begin{tabular}{cccccc}
\hline $J_{K_aK_c}'-J_{K_aK_c}''$ & experiment & Michaud et al. &
error (\%) & present & error (\%) \\
$1_{01}-0_{00}$ & 8117.3520   & 8257.2  & 1.7  & 8031.5   & -1.1       \\
$2_{12}-1_{11}$ & 15210.1951  & 15663.6 & 3.0  & 14938.0  & -1.8       \\
$2_{02}-1_{01}$ & 16196.1303  & 16485.6 & 1.8  & 16048.1  & -0.9       \\
$1_{10}-1_{01}$ & 16670.9616  & 18388.7 & 10.3 & 18366.9  & 10.2      \\
$2_{11}-1_{10}$ & 17097.2104  & 17341.8 & 1.4  & 16855.9  & -1.4     \\
$3_{13}-2_{12}$ & 22803.7703  & 23477.0 & 3.0  & 22400.0  & -1.8    \\
$1_{11}-0_{00}$ & 23843.1092  & 25806.4 & 8.2  & 25421.1  & 6.6     \\
$3_{03}-2_{02}$ & 24196.9788  & 24655.2 & 1.9  & 23987.6  & -0.9    \\
$3_{12}-2_{11}$ & 25625.2174  & 25994.1 & 1.4  & 25270.5  & -1.4     \\ \hline
\end{tabular}
\end{center}
\label{table:2}
\end{table*}%

In summary, as observed previously for smaller systems such as
CO-H$_2$ \citep{jankowski12} and HCN-H$_2$ \citep{denis13}, the
comparison between spectroscopic and theoretical data for
HC$_3$N--H$_2$ confirms that van der Waals rigid-rotor PES computed at
the CCSD(T) level with large basis sets can be used with confidence
for analyzing van der Waals complexes spectroscopy.

\section{Scattering calculations}

Scattering calculations for HC$_3$N-H$_2$ were performed using both
quantum and classical approaches. The objective was to obtain cross
sections for all transitions among the first 38 rotational levels of
HC$_3$N, i.e. up to $j_1=37$ which lies 213.3~cm$^{-1}$ above $j_1=0$
and just below the first excited bending mode $\nu_7$ which opens at
only 222~cm$^{-1}$. In practice, for levels higher than $j_1=30$, the
basis set necessary for converged quantum calculations would require
an excessive number of (open and closed) states, including
vibrationally excited states. In addition to the substantial CPU cost,
such calculations would require the determination of a flexible,
nonrigid-rotor, HC$_3$N--H$_2$ PES, which is beyond the scope of this
work. For the high-lying levels close to the vibrational threshold
($j_1=31-37$), the approximate but efficient classical approach was
therefore employed. Details on these scattering calculations are given
below.

\subsection{Quantum calculations}

The scattering quantum calculations were conducted at the full
coupled-channel level using the \texttt{MOLSCAT} program
\citep{molscat95}. The spherical harmonics expansion of the W07 PES,
including 97 angular functions, was employed and the coupled
differential equations were solved using the hybrid modified
log-derivative Airy propagator. Total energies up to 2120~cm$^{-1}$
were considered using a fine grid below 200~cm$^{-1}$ (with increments
as small as 0.01~cm$^{-1}$) for the correct assessment of resonances
close to thresholds. The total number of collision energies was 324
for para-H$_2$($j_2=0$) and 344 for ortho-H$_2$($j_2=1$). The highest
rotational level of HC$_3$N in the basis set was $j_1=44$ while one
single rotational level of H$_2$ was included for both the para
($j_2$=0) and ortho ($j_2$=1) modifications. The level $j_2$=2 of
para-H$_2$ was neglected because it was found to affect the cross
sections by less than 10-20\% on average \citep[see also][]{wernli07b}
while the CPU cost was increased by large factors. Thus, at a total
energy of 100~cm$^{-1}$, the number of channels was increased by a
factor of $\sim$6 (exceeding 4,000 channels) corresponding to a factor
of $\sim$ 200 in CPU time. We note that we performed a few high-energy
calculations with a basis set $j_2=0, 2$ in order to compare the cross
sections for H$_2$($j_2=2$) with those for H$_2$($j_2=1$), see
Section~4.3 below. The neglect of HC$_3$N levels $j_1>44$ was found to
change cross sections by less than a few percent for all transitions
among the first 31 levels of HC$_3$N. The maximum value of the total
angular momentum $J$ used in the calculations was $J$=130 at the
highest collisional energy. The rotational constants were taken as
$B_e$=60.853~cm$^{-1}$ for H$_2$\footnote{This value was used for
  consistency with our previous calculations. Using
  $B_0$=59.322~cm$^{-1}$ instead of $B_e$ would not modify the cross
  sections since a single rotational level of H$_2$ was included.} and
$B_0$=0.151739~cm$^{-1}$ for HC$_3$N, as in \cite{wernli07}. Finally,
convergence was checked as a function of the propagator step size
(parameter STEPS) and the other propogation parameters were taken as
the default \texttt{MOLSCAT} values.

Rate coefficients were obtained for kinetic temperatures in the range
10-300~K by integrating the product of the cross section by the
velocity over the Maxwell-Boltzmann distribution of velocities at each
temperature. Cross sections and rate coefficients are presented in
Section~4 below.

\subsection{Classical calculations}

The objective of the classical calculations was to provide rate
coefficients for all (classically allowed) transitions among the
levels $j_1=31-37$. Kinetic temperatures were restricted to the range
100-300~K since rate coefficients at lower temperatures are available
in \cite{wernli07}, where full details on the QCT approach can be
found. Briefly, it consists in solving the classical Hamilton
equations of motion instead of the Schr\"{o}dinger equation. In the
Monte-Carlo QCT approach, batches of trajectories are sampled with
random (Monte-Carlo) initial conditions and they are analyzed through
statistical methods. In the canonical formalism employed here, the
initial collision energies are selected according to the
Maxwell-Boltzmann distribution of velocities and rate coefficients,
instead of cross sections, are directly obtained (see Eqs.~(3) and (4)
in \cite{wernli07}). State-to-state rate coefficients are obtained by
use of the correspondence principle combined with the bin histogram
method for extracting the final $j_1$ quantum number.

Preliminary calculations had shown that QCT calculations for
para-H$_2$($j_2=0$) and ortho-H$_2$($j_2=1$) give very similar cross
sections \citep{wernli06}. As in \cite{wernli07}, we therefore
computed QCT cross sections for para-H$_2$($j_2=0$) only, using the
W07 PES averaged over H$_2$ orientation and fitted with a bicubic
spline function. The classical equations of motion were numerically
solved using an extrapolation Bulirsh-Stoer algorithm and numerical
derivatives for the potential. The total energy and total angular
momentum were checked and conserved up to seven digits, i.e. within
0.01~cm$^{-1}$ for the energy. The maximum impact parameter $b_{max}$
was found to range between 20 and 21~bohr for temperatures in the
range 100-300~K. Batches of 10,000 trajectories were run for each
initial level $j_1=31-37$ and each temperature, resulting in a total
of 770,000 trajectories. It should be noted that a particular
advantage of the QCT method is that the computational time decreases
with increasing collision energy, in contrast to quantum methods. QCT
calculations however ignores purely quantum effects such as
interference and tunneling, as will be shown below.

\subsection{Hyperfine excitation}

Due to the presence of a nitrogen atom, the HC$_3$N molecule possess a
hyperfine structure. Indeed, the coupling between the nuclear spin
($I=1$) of the nitrogen atom and the molecular rotation results in a
splitting of each rotational level $j_1$, into 3 hyperfine levels
(except for the $j_1=0$ level). Each hyperfine level is designated by
a quantum number $F_1$ ($F_1=I+j_1$) varying between $|I-j_1|$ and
$I+j_1$. As already discussed \citep[e.g.][and references
  therein]{faure12,lanza14}, if the hyperfine levels are assumed to be
degenerate, it is possible to simplify considerably the hyperfine
scattering problem. Almost exact hyperfine resolved cross sections can
then be obtained from nuclear spin-free $S$-matrices using the
so-called recoupling approach. This approach requires to store the
$S$-matrices and to compute hyperfine resolved $S$-matrices by
properly combining the nuclear spin-free $S$-matrices. However, in the
case of HC$_3$N--H$_2$ collisions, the recoupling approach becomes
rapidly prohibitive in terms of both memory and CPU time because of
the small rotational constant of the target molecule, i.e. the large
number of channels to include.

Alternatively, hyperfine resolved rate coefficients can be directly
obtained from rotational rate coefficients using the
scaled-infinite-order-sudden limit (S-IOS) method first introduced by
\cite{neufeld94} in the case of HCl-He collisions and then extended to
the hyperfine excitation of linear molecules by para- and ortho-H$_2$
by \cite{lanza14}. This adiabatic method is expected to be reliable at
high collision energies and/or for target molecules with small
rotational constants so that the rotational period is small compared
to the collision time scale. In practice, the hyperfine rate
coefficients (or cross sections) are obtained by scaling the
coupled-channel rotational rate coefficients by the ratio of hyperfine
and rotational IOS rate coefficients as follows:
\begin{equation} \label{formula_SIOS}
k^{S-IOS}_{j_1,F_1,j_2 \to j'_1F'_1,j'_2}=\frac{k^{IOS}_{j_1F_1,j_2
    \to j'_1F'_1,j'_2}} {k^{IOS}_{j_1,j_2 \to j'_1,j'_2}}
k^{CC}_{j_1,j_2 \to j'_1,j'_2},
\end{equation}
where $k^{IOS}_{j_1F_1,j_2 \to j'_1F'_1,j'_2}$ and $k^{IOS}_{j_1,j_2
  \to j'_1,j'_2}$ can be found in Eqs.~(6) and (7) of \cite{lanza14}.
We note that for quasi-elastic transitions (those with $j'_1=j_1$) no
scaling is applied, as explained in \cite{faure12}. We note also that
Eq.~(1) guarantees that the summed hyperfine rate coefficients are
identical to the coupled-channel pure rotational rate coefficients.

The S-IOS approach was applied recently to the HCl--H$_2$ collisional
system \citep{lanza14}. Despite the large rotational constant of HCl,
the S-IOS method was found to be accurate to within a factor of 2-3 in
the case of collisions with para-H$_2$($j_2 = 0$) at intermediate and
high kinetic energies ($>200$~cm$^{-1}$). It was however found to fail
by factors larger than 3 in the case of ortho-H$_2$($j_2 = 1$). As
HC$_3$N is much heavier than HCl and hence more adapted to IOS type
methods, one can expect the accuracy of the S-IOS approach to be
substantially better.



\section{Results}


\subsection{Collisional propensity rules}

\begin{figure}
\includegraphics*[width=7.5cm,angle=-90.]{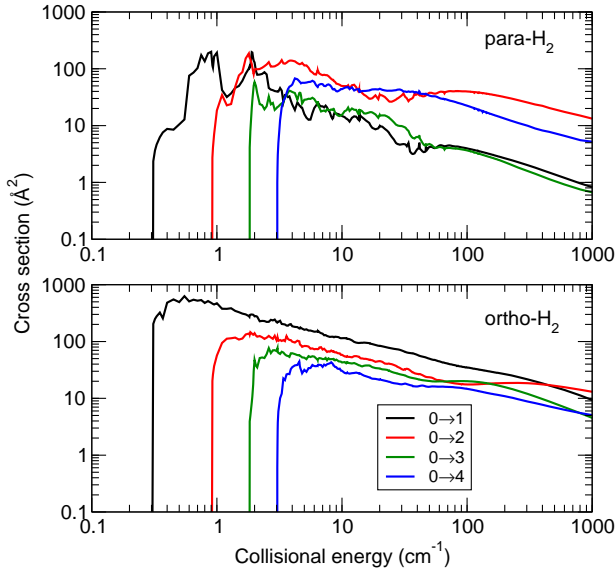}
\caption{Cross sections for rotational excitation out of the ground
  state of HC$_3$N $j_1=0$ into $j'_1=1, 2, 3$ and 4 as function of
  collisional energy. Upper and lower panels correspond to collisions
  due to para-H$_2$($j_2=0$) and ortho-H$_2$($j_2=1$), respectively.}
\label{xs}
\end{figure}

In Fig.\ref{xs}, cross sections for the rotational excitations
$j_1=0\rightarrow 1, 2, 3, 4$ are plotted as function of collision
energy for the colliders para-H$_2$($j_2=0$) and
ortho-H$_2$($j_2=1$). It is first noticed that these cross sections
show prominent resonances at low energies, especially in the case of
para-H$_2$($j_2=0$). These resonances, of both shape and Feshbach
types, arise from purely quantum effects. They have been observed
experimentally only recently \citep[see e.g.][in the case of
  CO--H$_2$]{chefdeville12,chefdeville15}. We also observe in the case
of para-H$_2$($j_2=0$) that transitions with even $\Delta j_1$
(i.e. $\Delta j_1=2, 4$) have the largest cross sections, with $\Delta
j_1=4$ being even dominant in the energy range
$\sim$15-45~cm$^{-1}$. This result was interpreted by \cite{wernli07}
as originating from the shape of the HC$_3$N--para-H$_2$ PES, being
nearly a prolate ellipsoid. It should be noted that these results are
unchanged when the basis set on H$_2$ include the $j_2=2$ level
\citep[see Fig.~1 in][]{wernli07b}. In contrast, in the case of
ortho-H$_2$, cross sections follow an energy-gap law behaviour with
$\Delta j_1=1>\Delta j_1=2 >\Delta j_1=3$, etc. We note, still, that
above $\sim$400~cm$^{-1}$, the transition $0\rightarrow 2$ dominates.

These results illustrate clearly the difference between para- and
ortho-H$_2$. It was previously observed for many other systems where
the target has a large dipole, e.g. for HCN--H$_2$ \citep{vera14}. In
the case of formaldehyde (H$_2$CO), it was even used to indirectly
constrain the ortho-to-para ratio of H$_2$
\citep{troscompt09}. Indeed, the general distinction between para- and
ortho-H$_2$ is attributable to the permanent quadrupole moment of
H$_2$, which vanishes for $j_2=0$ but not for $j_2>0$. When the dipole
of the target is significant, the long-range dipole-quadrupole
interaction term is large and the difference between
para-H$_2$($j_2=0$) and ortho-H$_2$($j_2=1$) is substantial (at the
quantum level).

\subsection{Comparison between classical and quantum rates}

\begin{figure}
\includegraphics*[width=7.5cm,angle=-90.]{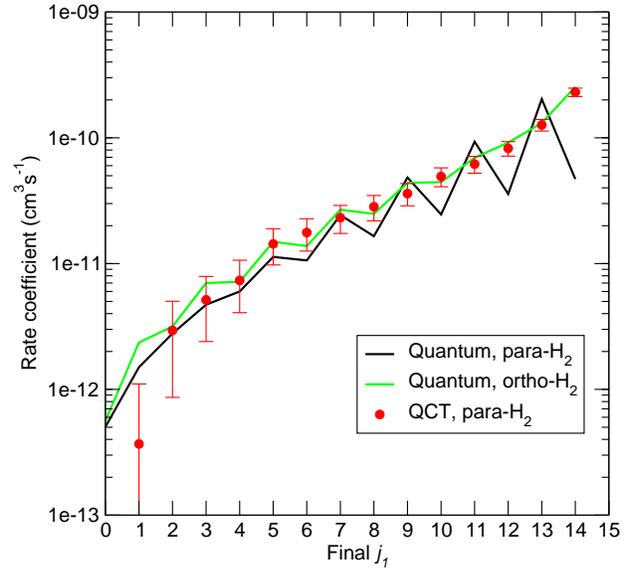}
\caption{Rate coefficients for the HC$_3$N deexcitation
  $j_1=15\rightarrow j'_1$ by H$_2$ at 100~K. Quantum coupled-channel
  results are given by the solid black and light green lines for
  para-H$_2$($j_2=0$) and ortho-H$_2$($j_2=1$), respectively. QCT
  results are represented by the solid circles with error bars
  representing two standard deviations.}
\label{qct}
\end{figure}

A comparison between quantum and classical rate coefficients is
presented in Fig.~\ref{qct} for HC$_3$N initially in level
$j_1=15$. The kinetic temperature is fixed at 100~K and the rate
coefficients are plotted as function of the final HC$_3$N level
$j'_1$. We first notice that the quantum results for para-H$_2(j_2=0$)
show a pronounced even $\Delta j_1$ (``near homonuclear'') propensity,
as expected from the cross sections reported above. This propensity is
not observed at the QCT level whereas the PES are identical,
demonstrating that this result indeed arises from a purely quantum
effect. In fact, this propensity was explained semi-classically by
\cite{mccurdy77} in terms of an interference effect related to the
even anisotropy of the PES. Interferences are of course ignored in QCT
calculations so that the ``zig-zags'' are totally absent. Now when
ortho-H$_2$($j=1$) is the projectile, the zig-zags are almost entirely
suppressed and, interestingly, the quantum results are in excellent
agreement with the QCT calculations, within error bars (except for the
largest $\Delta j_1=14$ transfer). This demonstrates that the
quadrupole moment of H$_2$, which vanishes for H$_2(j_2=0)$, plays a
crucial role by breaking the even symmetry of the PES and suppressing
the interference effect. But it does not modify the magnitude of the
cross sections, which are well reproduced by purely classical
mechanics with para-H$_2(j_2=0$). This result also suggests that the
scattering process is dominated by the short-range part of the PES,
i.e. by the ellipsoidal shape of the potential. It should be noted
that the interference structure is extremely sensitive to the PES
anisotropy and rotationally state-selected experiments which resolve
this structure would provide a critical test of theory \citep[see
  e.g.][in the case of CO+He]{carty04}.

These results show that QCT calculations can be employed with
confidence to mimic collisions with rotationally excited H$_2$($j_2\ge
1)$. They should be reliable to within 10-30\% for rate coefficients
larger than $\sim 10^{-12}$~cm$^3$s$^{-1}$. Smaller rate coefficients
have much larger uncertainties because they correspond to {\it rare}
or ``classically forbidden'' transitions with small probabilities.

\subsection{Thermalized rate coefficients}

The ortho-to-para ratio of H$_2$ can be out of thermal equilibrium in
the ISM \citep[e.g.][and references therein]{faure13}. It is thus
crucial to consider the two nuclear spin species of H$_2$ as two
separate colliders. We can however generally assume that each nuclear
spin species has a thermal distribution of rotational levels. At
temperatures below $\sim $80~K, only the ground states are
significantly populated so that in practice only para-H$_2$($j_2=0$)
and ortho-H$_2$($j_2=1$) must be considered. At higher temperatures,
however, the levels $j_2=2, 3$ and 4 also play an important role
because their (thermal) populations become larger than 1\% above
$\sim$80~K, $\sim$160~K and $\sim$260~K, respectively. Due to the
prohibitive CPU cost, we did not perform extensive scattering
calculations for these excited levels but a few energy points were
computed. We have thus found that rotational cross sections for H$_2$
initially in $j_2=2$ differ by less than $\sim$ 20-30\% with those for
H$_2$ in $j_2=1$. As a result, it can be assumed that all cross
section, and by extension all rate coefficients, for H$_2$($j_2>1$)
are identical to those for H$_2$($j_2=1$). This result was observed
previously for many other systems before, as a rather general rule
\cite[e.g.][and references therein]{daniel14}. We note that at higher
temperatures where the H$_2$ levels can be (de)excited, this rule
holds for ``effective'' rate coefficients which are summed over the
final H$_2$ levels.

For para-H$_2$, we have thus computed the ``thermalized'' rate
coefficients by weighting the para-H$_2$($j_2=0$) rate coefficients,
$k_{j_1, j_2=0\rightarrow j'_1}(T)$\footnote{Since rotational
  transitions in H$_2$ are neglected in our calculations, the
  calculated rate coefficients are ``effective'' and the final H$_2$
  level $j'_2$ can be omitted from the notation.}, by the thermal
distribution of the $j_2=0$ level and the ortho-H$_2$($j_2=1$) rate
coefficients, $k_{j_1, j_2=1\rightarrow j'_1}(T)$, by the thermal
distribution of all para levels with $j_2>0$:
\begin{equation}
k_{j_1, p\rm{H}_2\rightarrow j'_1}(T)=\rho_0k_{j_1,
  j_2=0\rightarrow j'_1}(T)+(1-\rho_0)k_{j_1, j_2=1\rightarrow
  j'_1}(T),
\end{equation}
where
\begin{equation}
\rho_0=\frac{1}{\sum_{j_2=0,2,4,...}(2j_2+1)\exp(-E_{j_2}/k_BT)}
\end{equation}
is the thermal population of $j_2=0$.

For ortho-H$_2$, all H$_2$ levels were assumed to have the same rate
coefficients so that the ``thermalized'' rate coefficients were simply
obtained as:
\begin{equation}
k_{j_1, o\rm{H}_2\rightarrow j'_1}(T)=k_{j_1, j_2=1\rightarrow j'_1}(T).
\end{equation}

In Fig.~3, the thermalized rate coefficients for para-H$_2$ and
ortho-H$_2$ are plotted as function of temperature for the
ground-state transition $j_1=1\rightarrow 0$ at 9.1~GHz. The
contribution of $j_2=2$ is significant above $\sim$80~K, as expected,
and it is here amplified by the fact that the corresponding rate
coefficient (set equal to $k_{j_1, j_2=1\rightarrow j'_1}(T)$) is
about a factor of 10 larger than $k_{j_1, j_2=0\rightarrow j'_1}(T)$
(see Eq.~2). At 300~K, the contribution of $j_2=2$ (and to much a
lesser extent $j_2=4$) increases the rate coefficient of para-H$_2$ by
a factor of $\sim 6$. It is therefore of crucial importance at these
temperatures.

\begin{figure}
\includegraphics*[width=7.5cm,angle=-90.]{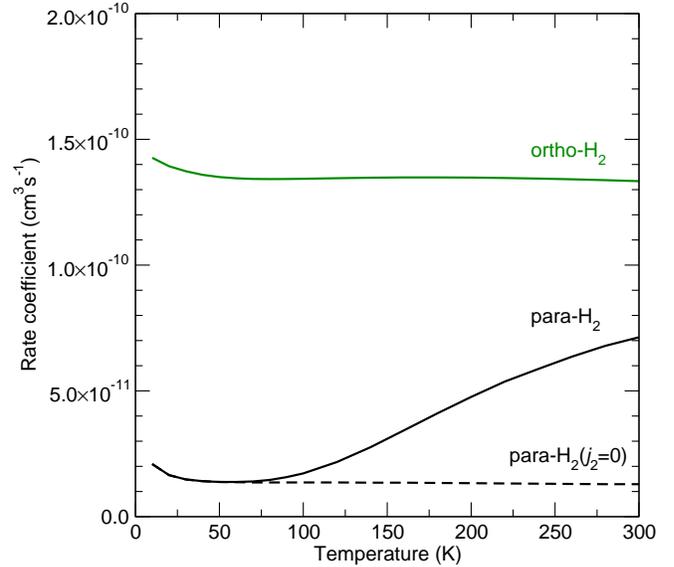}
\caption{Rate coefficients for the HC$_3$N deexcitation
  $j_1=1\rightarrow 0$ by para-H$_2$ and ortho-H$_2$ in the
  temperature range 10-300~K. The thermalized rate coefficient for
  para-H$_2$($j_2=0, 2, 4, ...$) is compared to the contribution of
  the ground state $j_2=0$.}
\label{para}
\end{figure}

The whole set of thermalized rate coefficients for the lowest 38
rotational levels of HC$_3$N and kinetic temperatures in the range
10-300~K are available at the
\texttt{LAMDA}\footnote{http://www.strw.leidenuniv.nl/$\sim$moldata}
\citep{schoier05} and
\texttt{BASECOL}\footnote{http://basecol.obspm.fr} \citep{dubernet13}
data bases. We note that for HC$_3$N levels between $j_1=31$ and
$j_1=37$, only QCT rate coefficients for para-H$_2$($j_2=0$) are
available and no thermal averaging was applied to this set which is
employed for both para-H$_2$ and ortho-H$_2$. 

\subsection{Hyperfine cross sections and rate coefficients}

\begin{figure*}
\includegraphics*[width=6.5cm,angle=0.]{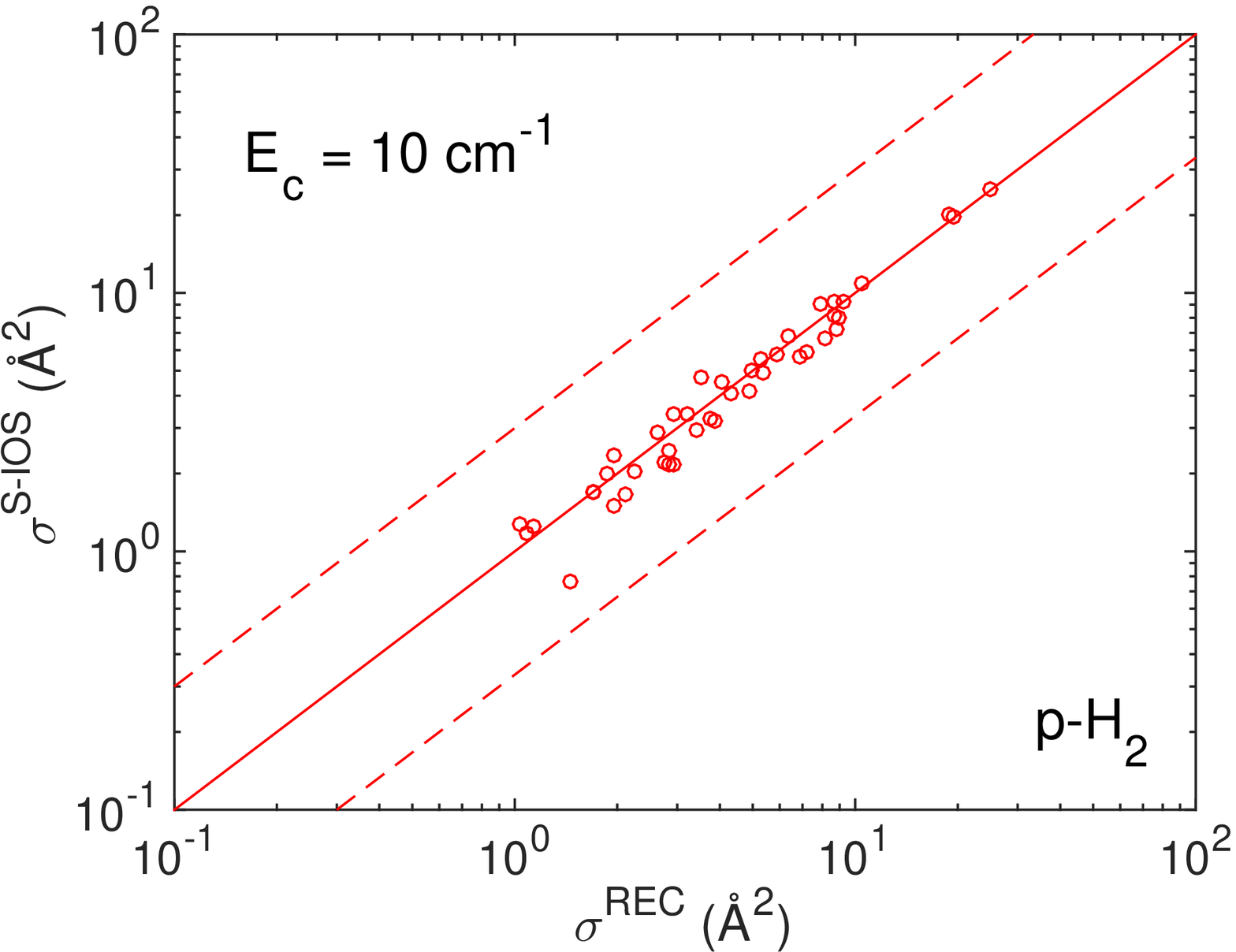}
\includegraphics*[width=6.5cm,angle=0.]{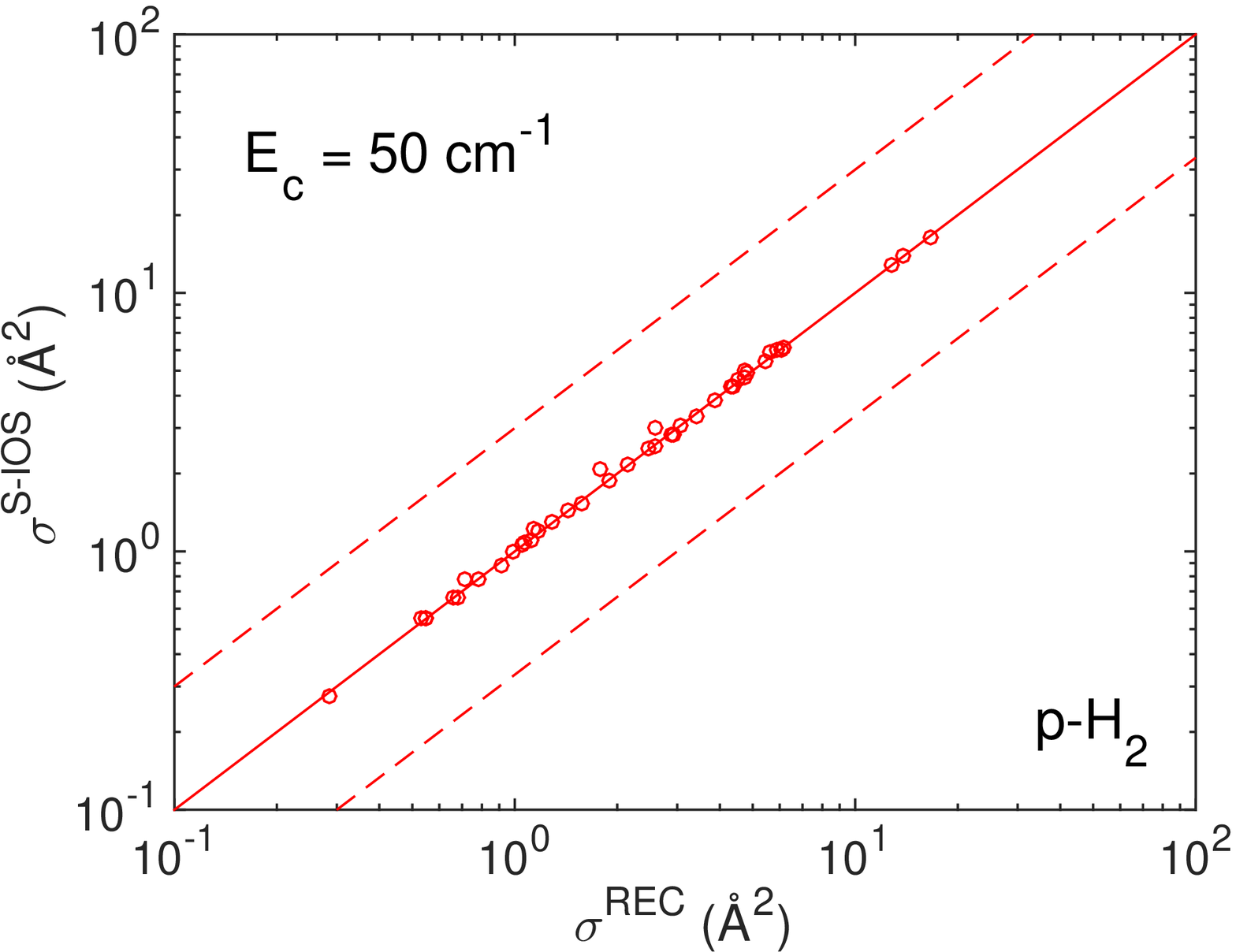}
\includegraphics*[width=6.5cm,angle=0.]{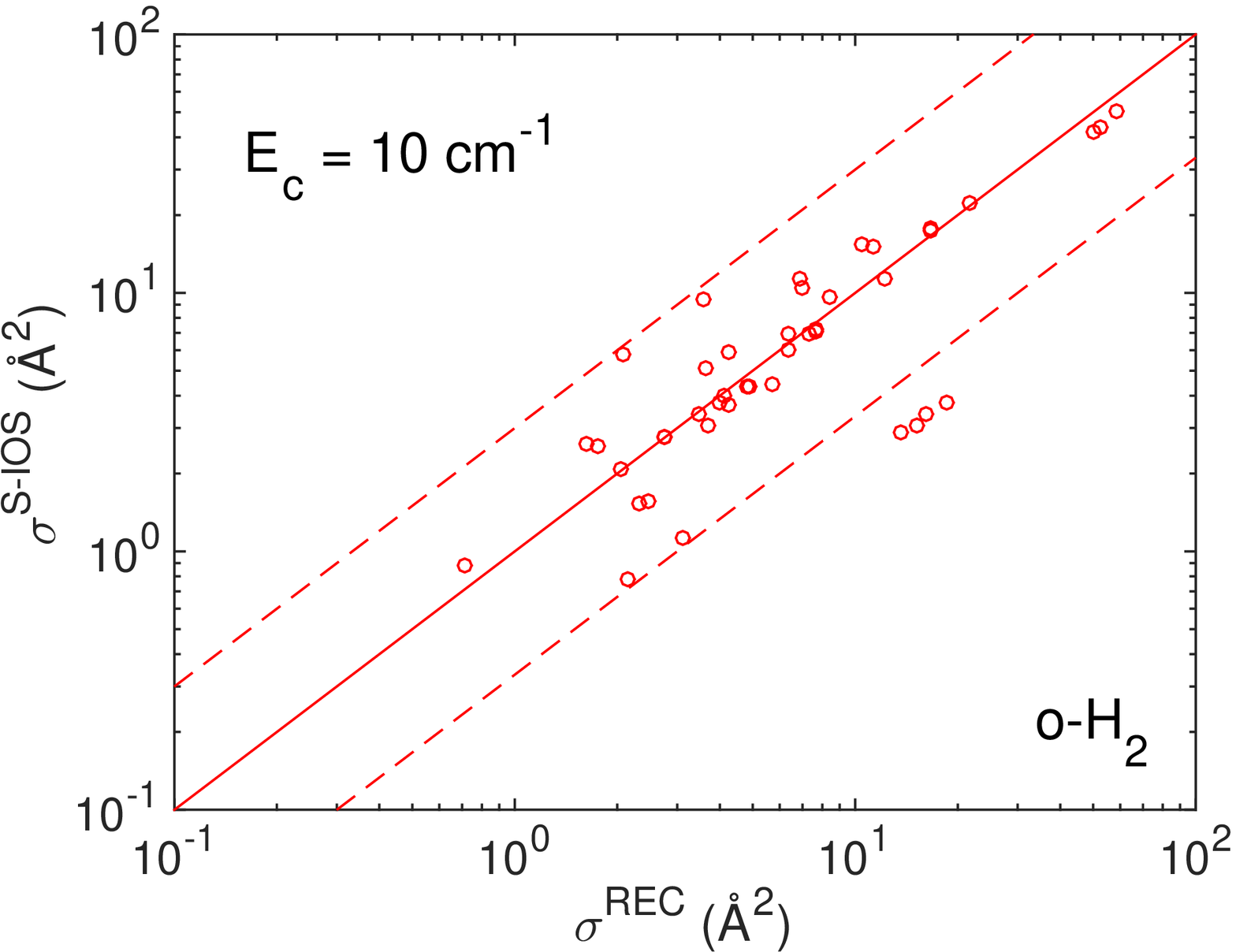}
\includegraphics*[width=6.5cm,angle=0.]{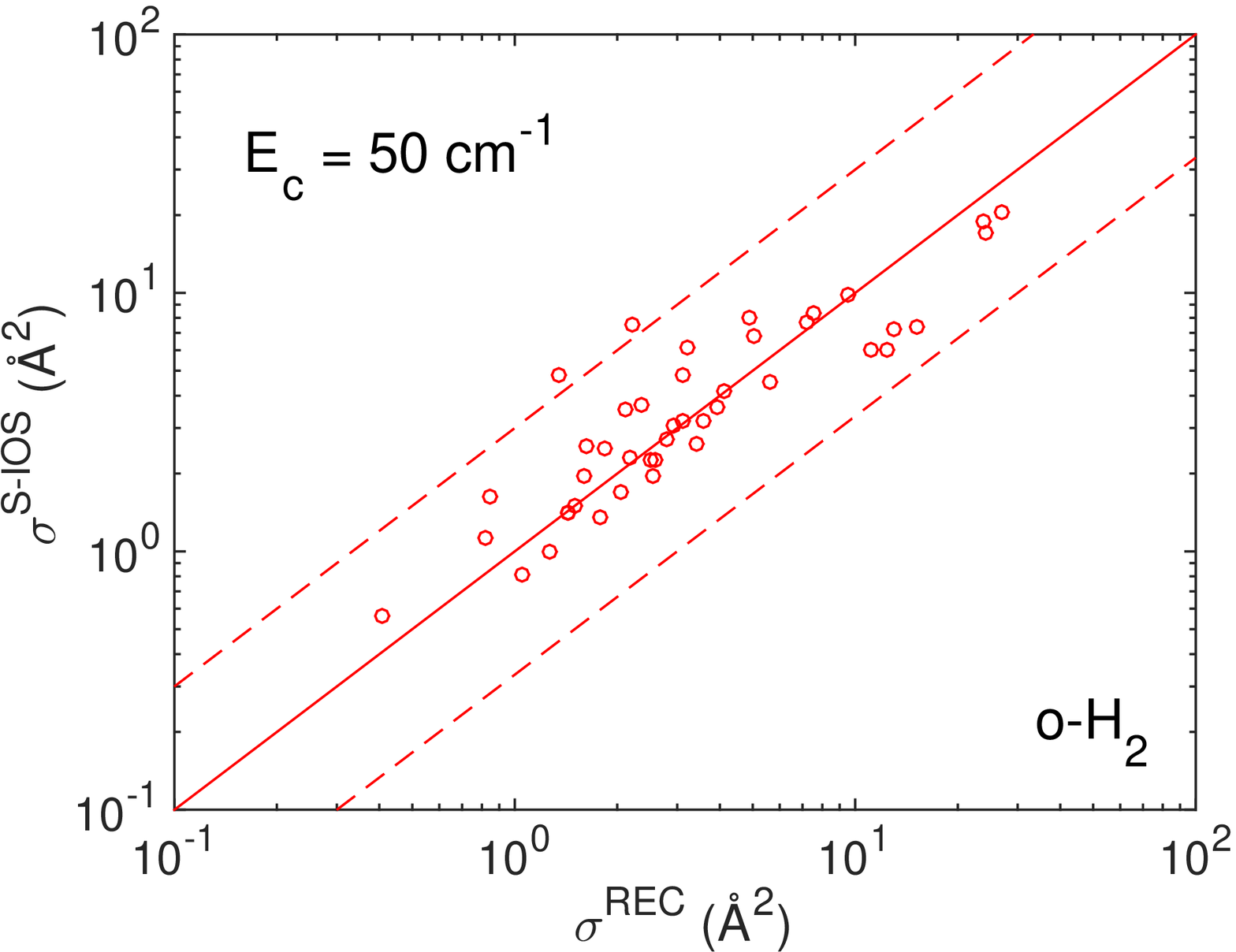}
\caption{Comparison between HC$_3$N--H$_2$ recoupling and S-IOS
  hyperfine cross sections for all the de-excitation transitions from
  $j_1$=5 at two different collisional energies. The vertical axis
  represents the hyperfine S-IOS cross sections and the horizontal
  axis represents the corresponding hyperfine recoupling cross
  sections. The two dashed lines in each panel delimit the region
  where the cross sections differ by less than a factor of 3.}
\label{Comp_hyp_CC_SIOS}
\end{figure*}

In order to evaluate the accuracy of the S-IOS method (see
Section~3.3) in the case of HC$_3$N--H$_2$ collisions, we have
computed recoupling and S-IOS hyperfine cross sections at two selected
collisional energies (10 and 50~cm$^{-1}$). In the above
Eq.~(\ref{formula_SIOS}), cross sections were employed instead of rate
coefficients. Fig.~\ref{Comp_hyp_CC_SIOS} shows a comparison between
recoupling and S-IOS hyperfine cross sections for all the
de-excitation transitions from the initial level $j_1$=5, including
quasi-elastic transitions (those with $j_1=j_1'$ and $F_1 \ne F_1'$),
for collisions with para-H$_2(j_2=0)$ and ortho-H$_2(j_2=1)$. We note
that the S-IOS approach imposes selection rules (through Wigner $6j$
symbols) and some cross sections (not plotted) are strictly zero. The
corresponding transitions are those between the ($j_1=1, F_1=0$) level
and levels with $F_1=j_1$, e.g. $(1, 0)\rightarrow (1, 1)$. At the
close-coupling level, these transitions have indeed cross sections
lower than $\sim$1~\AA$^2$.

For collisions with para-H$_2(j_2=0)$, the agreement between
recoupling and S-IOS calculations is very satisfactory. At low
energies, the differences between the two sets of data are less than
20-30\% (especially for the dominant cross sections). At kinetic
energies greater than 50~cm$^{-1}$, cross sections are almost
identical. We conclude that the S-IOS approach can be used with
confidence to compute hyperfine rate coefficients in the case of
collisions with para-H$_2(j_2=0)$. For collisions with
ortho-H$_2(j_2=1)$, the agreement between recoupling and S-IOS
calculations is less satisfactory. At low energies, the two sets of
data agree within a typical factor of 3. At kinetic energies greater
than 50 cm$^{-1}$, the differences between the two sets decrease and
the average difference is lower than a factor of two.

To summarize, the S-IOS method can provide hyperfine resolved rate
coefficients with an average accuracy better than 20-30\% in the case
of collisions with para-H$_2$($j_2=0$) and within a factor of 2-3 in
the case of collisions with ortho-H$_2$($j_2=1$). With respect to the
memory and CPU cost of full recoupling calculations, the S-IOS
approximation therefore represents a suitable (and low-cost)
alternative for this system. Finally, in terms of radiative transfer
application, it should be noted that at moderate and high opacities,
where the relative hyperfine populations can significantly depart from
the statistical weights, the S-IOS method is notably better than the
statistical approach \citep[see][]{faure12}.

In practice, hyperfine resolved rate coefficients were obtained for
the lowest 61 hyperfine levels of HC$_3$N, i.e. up to $(j_1=20,
F_1=20$) which lies 63.73~cm$^{-1}$ above (0, 1), and for kinetic
temperatures in the range 10-100~K. This set of data is available at
the \texttt{LAMDA} and \texttt{BASECOL} data bases.

\section{Conclusion}

We have reported in this paper rate coefficients for the rotational
excitation of HC$_3$N by para- and ortho-H$_2$. The lowest 38
rotational levels of HC$_3$N were included and kinetic temperatures up
to 300~K were considered. The scattering calculations were performed
at the quasi-classical and quantum coupled-channel level using the
interaction potential of \cite{wernli07}. This potential was also
employed to compute the bound-states of the complex in order to make
comparisons with the spectroscopy measurements of Michaud et
al. (2011). Theory and experiment were found to agree within 0.5\% for
para-H$_2$($j_2=0$) and within a few percent for ortho-H$_2$($j_2=1$),
demonstrating the high accuracy of the potential. It appears from
these comparisons that the calculated state-to-state rotational rate
coefficients are likely to be accurate to about 20-30\%. Hyperfine
resolved rate coefficients were also deduced using the S-IOS
approximation, with a somewhat lower accuracy. The whole set of data
represent a significant improvement and extension over the previous
data of \cite{green78} and \cite{wernli07}.

The next step is to determine a flexible potential energy surface in
order to treat the ro-vibrational excitation of HC$_3$N. The lowest
vibrational state is the $\nu_7$ bending mode which lies at
222~cm$^{-1}$ above the ground vibrational state. Ro-vibrational
excitation due to collisions is therefore expected to play a role
above $\sim$300~K. Such calculations are highly challenging due to
excessively large number of channels involved. This problem has been
however tackled recently at the quantum coupled-channel level using
the rigid-bender approximation \citep{stoecklin13,stoecklin15}. This
latter approximation was also employed previously in quasi-classical
trajectories \citep{faure05} which offer an (economical) alternative
to quantum computations, as shown in the present work.

In summary, the present set of data, possibly complemented by
electron-impact rate coefficients \citep[as given in ][]{gratier13},
should help in modelling non-LTE HC$_3$N spectra in cold to warm
regions of the ISM. We note in particular that HC$_3$N has recently
been detected in protopolanetary disks
\citep{chapillon12,oberg15}. The collisional data provided here should
prove very useful in interpreting such observations.

\section*{Acknowledgements}

This research was supported by the CNRS national program 'Physique et
Chimie du Milieu Interstellaire'. Most of the computations presented
in this paper were performed using the CIMENT infrastructure
(https://ciment.ujf-grenoble.fr), which is supported by the
Rh\^one-Alpes region (GRANT CPER07 13 CIRA).

\bibliographystyle{mn2e}


\bsp

\label{lastpage}

\end{document}